\documentclass[12pt]{article}
\usepackage{graphicx}
\usepackage{units}
\usepackage{url}
\usepackage{amsmath}
\usepackage{amssymb}
\usepackage{amsfonts}
\usepackage{bm}
\usepackage{units}
\usepackage{textcomp}

\newcommand{\ee}{e^+e^-}
\newcommand{\dzero}{D^0}
\newcommand{\dbarzero}{\overline{D}{}^{0}}

\newcommand{\gev}{\,\unit{GeV}}

\newcommand{\br}[1]{\mathcal{B}_{#1}}
\newcommand{\srd}{r}
\newcommand{\kpi}{{K^-\pi^+}}

\newcommand{\strph}{\delta_{K\pi}}
\newcommand{\ycp}{y_\mathrm{CP}}
\newcommand{\mbc}{M_\mathrm{BC}}

\def\pbnr{}
\def\speaker{Xiao-Rui Lu}
\def\onbehalfof{the BESIII Collaboration}
\def\title{Measurements of strong phase in $D^0\to K\pi$ decay and $y_\mathrm{CP}$ at BESIII}
\def\affiliation{School of Physics\\
University of Chinese Academy of Sciences, 100049, Beijing, China}
\def\support{E-mail: xiaorui@ucas.ac.cn}

\newcommand\pubnumber{\pbnr}
\newcommand\pubdate{\today}
\def\Title#1{\begin{center} {\Large #1 } \end{center}}
\def\Author#1{\begin{center}{ \sc #1} \end{center}}

\newcommand{\OnBehalf}[1]{\sbox0{#1}\ifdim\wd0=0pt
        {}
	\else
	{\\on behalf of #1}
	\fi}
\newcommand{\SupportedBy}[1]{\sbox0{#1}\ifdim\wd0=0pt
        {}
	\else
	{\footnote{#1}}
	\fi}
\def\Address#1{\begin{center}{ \it #1} \end{center}}

\newcommand\pubblock{\includegraphics[width=5cm]{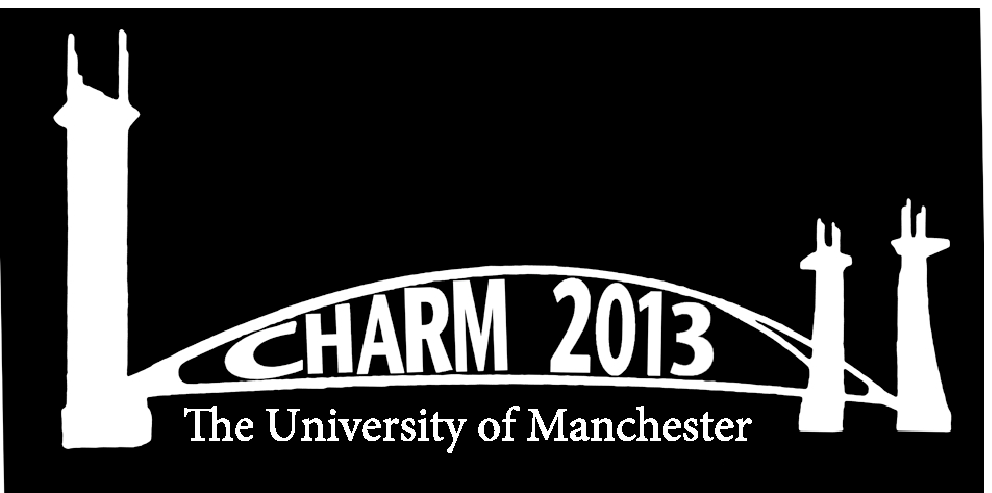}\hfill{\begin{tabular}{l} \pubnumber\\
         \pubdate  \end{tabular}}}
\newenvironment{Abstract}{\begin{quotation}  }{\end{quotation}}
\newenvironment{Presented}{\begin{quotation} \begin{center} 
             PRESENTED AT\end{center}\bigskip 
      \begin{center}\begin{large}}{\end{large}\end{center} \end{quotation}}
\def\Acknowledgements{\bigskip  \bigskip \begin{center} \begin{large}
             \bf ACKNOWLEDGEMENTS \end{large}\end{center}}
\def\venue{The 6$^{th}$ International Workshop on Charm Physics\\
(CHARM 2013)\\
Manchester, UK,  31 August -- 4 September, 2013}

\def\beq{\begin{equation}}
\def\eeq#1{\label{#1}\end{equation}}
\def\eeqn{\end{equation}}

\def\beqa{\begin{eqnarray}}
\def\eeqa#1{\label{#1}\end{eqnarray}}
\def\eeqan{\end{eqnarray}}

\let\bar=\overbar

\def\Dslash{\not{\hbox{\kern-4pt $D$}}}
\def\dslash{\not{\hbox{\kern-2pt $\del$}}}

\def\ee{e^+e^-}

\def\msb{{\bar{\ssstyle M \kern -1pt S}}}

\begin{document}
\begin{titlepage}
\pubblock

\vfill
\Title{\title}
\vfill
\Author{\speaker\SupportedBy{\support}\OnBehalf{\onbehalfof}}
\Address{\affiliation}
\vfill
\begin{Abstract}
 In this paper, I report the preliminary results of the strong phase difference $\cos\delta_{K\pi}$ between the doubly Cabibbo-suppressed process $\overline{D}{}^0\to K^- \pi^+$ and Cabibbo-favored $D^0\to K^- \pi^+$ at BESIII. In addition, the preliminary results of the  $D^0$-$\overline{D}{}^0$ mixing parameter $y_\mathrm{CP}$ by analyzing $CP$-tagged semileptonic $D$ decays are presented. These measurements were carried out based on the quantum-correlated technique in studying the process of $D^0\overline{D}{}^0$ pair productions of 2.92 fb$^{-1}$ $e^+e^-$ collision data collected with the BESIII detector at $\sqrt{s}$ = 3.773 GeV.
\end{Abstract}
\vfill
\begin{Presented}
\venue
\end{Presented}
\vfill
\end{titlepage}
\def\thefootnote{\fnsymbol{footnote}}
\setcounter{footnote}{0}
%

\section{Introduction}

 $\dzero$-$\dbarzero$ mixing originated from the short distance effect is highly suppressed by the the GIM mechanism~\cite{GIM} and by the CKM matrix elements~\cite{CKM} within standard model. However, long distance effect, which is not calculated reliable,  can manifest size of the mixing. Thus, to probe the $\dzero$-$\dbarzero$ mixing constitutes identifying the size of the long distance effect and searching for new physics~\cite{Bigi}. In addition, improving constraints on charm mixing is important for studying $CP$ violation ($CPV$) in charm physics.

Charm mixing is described by two dimensionless parameters
$$    x=2\frac{M_1-M_2}{\Gamma_1+\Gamma_2}\label{1} \hspace{1cm}    y=\frac{\Gamma_1-\Gamma_2}{\Gamma_1+\Gamma_2},$$
where $M_{1,2}$ and $\Gamma_{1,2}$ are the masses and widths of two mass eigenstates.
Because $CPV$ in $D$ decays is quite small compared with the mixing parameters,
it is reasonable to assume no direct $CPV$. The parameter $\ycp$ is defined as follows~\cite{Xing:1996pn}
$$\ycp\equiv\frac{1}{2}[y{\rm{cos}}\phi(|\frac{q}{p}|+|\frac{p}{q}|) - x {\rm{sin}}\phi  (|\frac{q}{p}|-|\frac{p}{q}|)],$$
where $|\frac{p}{q}|$ and $\phi$ correspond to the $CPV$ in mixing and in interference between mixing and decay, respectively~\cite{PDG}.
So far, $\ycp$ is measured mainly from the time-dependent analysis of $\dzero\to K^+K^-$ and $\dzero\to \pi^+\pi^-$.
At BESIII, $\ycp$ can be measured by comparing decay rates of $D$ semileptonic decays in different $CP$-eigenstates. BESIII results provide further constrains to the world measurements of the mixing parameters in $D$ sector. In case no $CPV$, we have $|\frac{p}{q}|=1$ and $\sin\phi=0$. Hence, $\ycp=y$.

In the context of $CP$ conservation, the mass eigenstates $D_1$  and $D_2$ can be written as
\begin{equation}
    |D_1\rangle\equiv\frac{|D^0\rangle+|\dbarzero\rangle}{\sqrt{2}}, \hspace{30pt}
    |D_2\rangle\equiv\frac{|D^0\rangle-|\dbarzero\rangle}{\sqrt{2}}.\label{4} \nonumber
    \end{equation}
If we take the phase convention $CP|D^0\rangle=+|\dbarzero\rangle$~\cite{PDG}, $D_1$ and $D_2$ are also $CP$ eigenstates of $CP$-even and $CP$-odd, respectively. The strong phase difference $\strph$ between the doubly Cabibbo-suppressed (DCS) decay $D^0\to K^+\pi^-$ and the corresponding Cabibbo-favored (CF) $\dbarzero\to K^-\pi^+$ is denoted as
    \begin{equation}
    \frac{\langle K^-\pi^+|\dbarzero\rangle}{\langle K^-\pi^+|\dzero\rangle}
    = -\srd e^{-i\strph},\label{5}  \nonumber
    \end{equation}
which plays an important role in precise determinations of $\dzero$-$\dbarzero$ mixing parameters.
Here
    \begin{equation}
    \srd=\left|\frac{\langle K^-\pi^+|\dbarzero\rangle}{\langle K^-\pi^+|\dzero\rangle}\right|.\label{6}  \nonumber
    \end{equation}
In the limit of $CP$ conservation, we have
    \begin{equation}
    \langle K^-\pi^+|\dbarzero\rangle=\langle K^+\pi^-|\dzero\rangle, \hspace{30pt}
     \langle K^-\pi^+|\dzero\rangle=\langle K^+\pi^-|\dbarzero\rangle.  \nonumber
    \end{equation}
Hence, $\strph$ is the same in the final states of $K^-\pi^+$ and $K^+\pi^-$. In this paper, we use the notation of $K^-\pi^+$, and its charge conjugation mode is always implied to be included.

The most precise determination of the size of the mixing comes from the measurement of the time dependence of the decay rate of the wrong-sign process $D^0\to K^+\pi^-$. These analyses are sensitive to $y'\equiv y\cos\strph-x\sin\strph$ and $x'\equiv x\cos\strph +y\sin\strph$~\cite{yp1}. The measurement of $\strph$ can allow $x$ and $y$ to be extracted from $x'$ and $y'$. An improved determination of $\strph$ is important for this extraction. Furthermore, finer precision of $\strph$ helps the $\gamma/\phi_3$ angle measurement in CKM matrix according to the so-called ADS method~\cite{PDG}.

Using the quantum-correlated technique, $\strph$ and $y_\mathrm{CP}$ can be measured in the mass-threshold production process $\ee\to\dzero\dbarzero$~\cite{Cheng:2007uj}. In this process, the initial system has $J^{PC}=1^{--}$; as a result, the $\dzero$ and $\dbarzero$ are in a $CP$-odd  quantum-coherent state.  At any time, the $D^0$ and $\dbarzero$ mesons are in opposite $CP$-eigenstates, until one of them decays~\cite{Bigi}. This provides an unique way to probe $D^0$-$\dbarzero$ mixing as well as the strong phases difference between $D^0$ and $\dbarzero$ decay amplitudes, taking advantage of the quantum coherence of $D^0$-$\dbarzero$ pairs.

In this paper, we present the preliminary results of $\strph$ and $\ycp$ that uses the quantum correlated productions of $\dzero$-$\dbarzero$ mesons at $\sqrt{s}=3.773\gev$ in $e^+e^-$ collisions with an integrated luminosity of 2.92\,fb$^{-1}$ collected with the BESIII detector~\cite{:2009vd}. Details of the BESIII detector can be found in Ref.~\cite{:2009vd}.

\section{Measurement of the strong phase difference $\strph$}

The strong phase difference $\strph$ can be accessed using the following formula
\begin{eqnarray}
   2\srd \cos\strph + y &=& (1+R_{\mathrm{WS}})\cdot\mathcal{A}_{CP\to K\pi},
    \label{11}
\end{eqnarray}
where $R_\mathrm{WS}$ is the decay rate ratio of the wrong sign process $\dbarzero\to\kpi$ and the right sign process $\dzero\to\kpi$~\cite{Asner:2008ft} and $\mathcal{A}_{CP\to K\pi}$ is the asymmetry between $CP$-odd and $CP$-even states decaying to $\kpi$
\begin{eqnarray}
\mathcal{A}_{CP\to K\pi}=\frac{\br{D_2\to \kpi}-\br{D_1\to \kpi}}{\br{D_2\to \kpi}+\br{D_1\to \kpi}}.\label{12}
\end{eqnarray}
Using $D$ tagging method in the quantum-coherent $\dzero$ pair production, we can calculate the branching fractions with
\begin{equation}
    \br{D^{CP\pm}\to K\pi} = \frac{n_{K\pi,CP\pm}}{n_{CP\pm}}\cdot\frac{\varepsilon_{CP\pm}}{\varepsilon_{K\pi,CP\pm}}.\label{22}
\end{equation}
Here, $n_{CP\pm}$ ($n_{K\pi,CP\pm}$) and $\varepsilon_{CP\pm}$ ($\varepsilon_{K\pi,CP\pm}$) are yields and detection efficiencies of single tags (ST) of $D\to CP\pm$ (double tags (DT) of $D\to CP\pm$, $\bar{D}\to K\pi$), respectively.
With external inputs of the parameters of $r$, $y$ and $R_{\mathrm{WS}}$,  we can extract $\strph$ from
$\mathcal{A}_{CP\to K\pi}$. Based on a dataset of 818\,pb$^{-1}$ of collision data collected with the
CLEO-c detector at the center of mass $\sqrt{s}=3.77\gev$, the CLEO collaboration
measured $\cos\strph=0.81^{+0.22+0.07}_{-0.18-0.05}$~\cite{CLEO-c2}. Using a global fit method with inclusion of the external mixing parameters, CLEO obtained $\cos\strph=1.15^{+0.19+0.00}_{-0.17-0.08}$~\cite{CLEO-c2}.

\begin{figure*}[hp]
\centering
\includegraphics[width=0.8\linewidth]{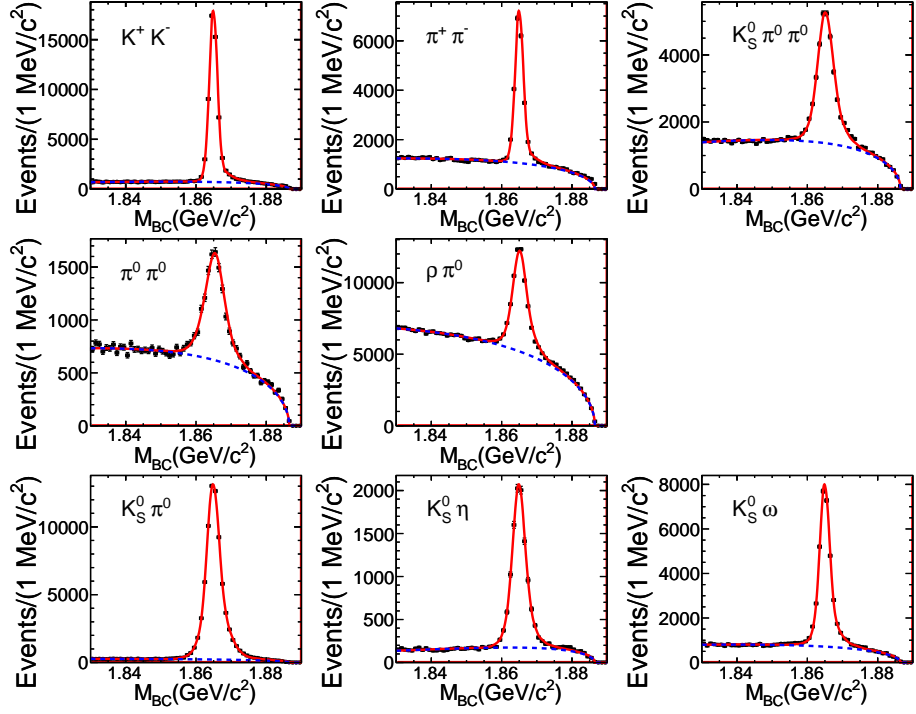}
\caption{ST $\mbc$ distributions of the $D\to CP\pm$ decays and fits to data. Data are shown in points with error bars. The solid lines show the total fits and the dashed lines show the background shapes. }
\label{fig:ST_all}
\end{figure*}

\begin{figure*}[hp]
\begin{center}
\includegraphics[width=0.8\linewidth]{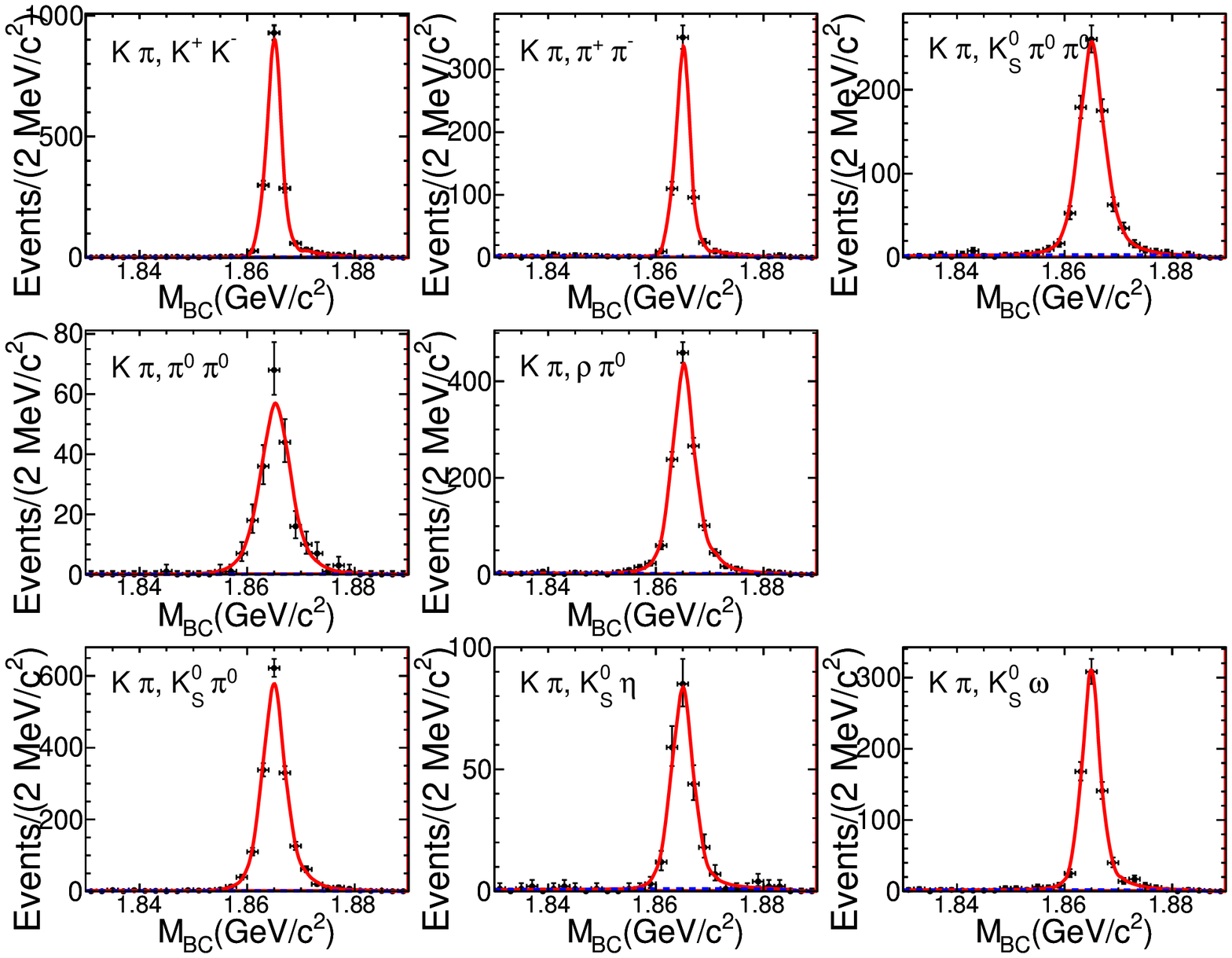}
\caption{DT $\mbc$ distributions and the corresponding fits. Data are shown in points with error bars. The solid lines show the total fits and the dashed lines show the background shapes.}{\label{fig:DT_all}}
\end{center}
\end{figure*}

We choose 5 $CP$-even $\dzero$ decay modes and 3 $CP$-odd modes, as listed in Tab.~\ref{tab:CP_Mode}, with $\pi^0\to\gamma\gamma$, $\eta\to\gamma\gamma$, $K^0_S\to\pi^+\pi^-$ and $\omega\to\pi^+\pi^-\pi^0$. Variable $$\mbc\equiv\sqrt{E^2_0/c^4-|\vec{p}_\mathrm{D}|^2/c^2}$$ is plotted in Fig.~\ref{fig:ST_all} to identify the $CP$ ST signals, where $\vec{p}_\mathrm{D}$ is the total momentum of the $\dzero$ candidate and $E_0$ is the beam energy. Yields of the $CP$ ST signals are estimated by maximum likelihood fits to data, in which signal shapes are derived from MC simulation convoluted with a smearing Gaussian function, and background functions are modeled with the ARGUS function~\cite{Argus}. In the events of the $CP$ ST modes, we reconstruct the $K\pi$ combinations using the remaining charged tracks with respect to the ST $D$ candidates. Similar fits are implemented to the distributions of $\mbc(D\to CP\pm)$ in the survived DT events to estimate yields of DT signals. The fits are shown in Fig.~\ref{fig:DT_all}.

\begin{table}[t!]
\begin{center}
\begin{tabular}{lc }
\hline
\hline
Type        & Mode  \\
\hline
Flavored    & $K^-\pi^+, K^+\pi^-$  \\
$CP+$         & $K^+K^-, \pi^+\pi^-, K^0_S\pi^0\pi^0, \pi^0\pi^0, \rho^0\pi^0$  \\
$CP-$         & $K^0_S\pi^0, K^0_S\eta, K^0_S\omega$  \\
\hline
\hline
\end{tabular}
\caption{$D$ decay modes reconstructed in the analysis of $\strph$.}
\label{tab:CP_Mode}
\end{center}
\end{table}

We get the asymmetry to be
\[
    \mathcal{A}_{CP\to K\pi}=(12.77 \pm 1.31^{+0.33}_{-0.31})\%,
\]
where the first uncertainty is statistical and the second is systematic. To measure the strong phase $\strph$ in Eq.~\eqref{11}, we quote the external inputs of $R_\mathrm{D}=r^2=3.47\pm 0.06$\textperthousand, $y=6.6\pm0.9$\textperthousand, and $R_\mathrm{WS}=3.80\pm0.05$\textperthousand ~ from HFAG 2013~\cite{HFAG} and PDG~\cite{PDG}. Hence, we obtain
\[
    \cos\strph = 1.03\pm0.12\pm0.04\pm0.01,
\]
where the first uncertainty is statistical, the second uncertainty is systematic, and the third uncertainty is due to the errors introduced by the external input parameters. This result is more precise than CLEO's measurement and provides the world best constrain to $\strph$.

\section{Measurement of $\ycp$}

For $D$ decays to any $CP$-eigenstate final states, their decay rates can be formulated to be
$$ R(D^{0}/\bar{D}^{0} \rightarrow CP\pm) \propto |A_{CP^{\pm}}|^2(1\mp y_{CP}).$$
When the partner $\bar{D}$ decays semileptonically in the $D\bar{D}$ pair production at threshold, double decay rates will be
$$R_{l;CP\pm} = |A_{l}|^{2}|A_{CP^{\pm}}|^{2}.$$
If we take the ratio of the single decay rate and the double decay rate and neglect higher order of $\ycp^2$, we can extract $\ycp$ using the following equation~\cite{Xing:1996pn}
$$ y_{CP} = \frac {1}{4}(\frac{R_{l;CP+}R_{CP-}}{R_{l;CP-}R_{CP+}}-\frac{R_{l;CP-}R_{CP+}}{R_{l;CP+}R_{CP-}} ).$$
Similar to the notations in Eq.~\eqref{22}, experimentally we denote the decay rate ratios of $\frac{R_{l;CP\pm}}{R_{CP\pm}}$ to be $B_{\pm}$ and
determine it with the $D$ tagging method
$$B_{\pm}= \frac{n_{l;CP\pm} }{n_{CP\pm}}\cdot\frac{ \varepsilon_{CP\pm}}{ \varepsilon_{l;CP\pm}}.$$
Hence, $y_{CP}=\frac{1}{4}[\frac{\tilde{B}_{+}}{\tilde{B}_{-}}-\frac{\tilde{B}_{-}}{\tilde{B}_{+}}]$, where
$\tilde{B}_{\pm}$ is combinations of different $CP$-tag mode $\alpha$ using the least square method
$$\chi^{2}= \sum_\alpha \frac{(\tilde{B}_{\pm}-B^{\alpha}_{\pm})^{2}}{(\sigma^{\alpha}_{\pm})^{2}}$$.

\begin{table}[t]
\centering
\begin{tabular}{ll}  \hline\hline
 Type  &   Modes \\ \hline
$CP^{+}$  & $K^{+}K^{-}$, $\pi^{+}\pi^{-}$, $K_{S}\pi^{0}\pi^{0}$ \\ \hline
$CP^{-}$  & $K_{S}^{0}\pi^{0}$, $K_{S}^{0}\omega$, $K_{S}^{0}\eta$ \\ \hline
$l^{\pm}$ & $Ke\nu$, $K\mu\nu$  \\ \hline \hline
\end{tabular}
\caption{$CP$-tag modes and $D$ semileptonic decay modes.}
\label{table:modes}
\end{table}

\begin{figure}[h]
\begin{center}
\includegraphics[width=4.cm]{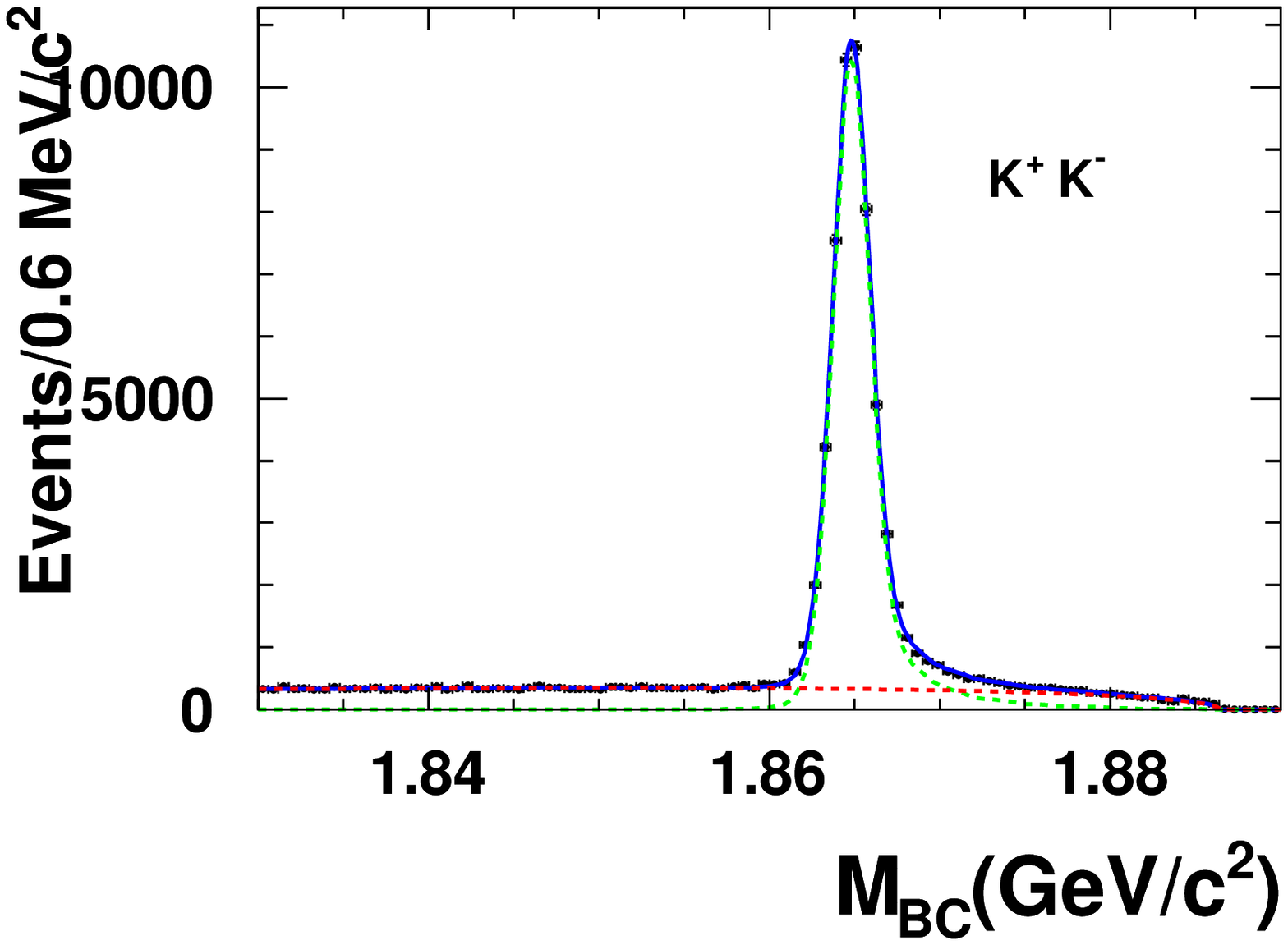}
\includegraphics[width=4.cm]{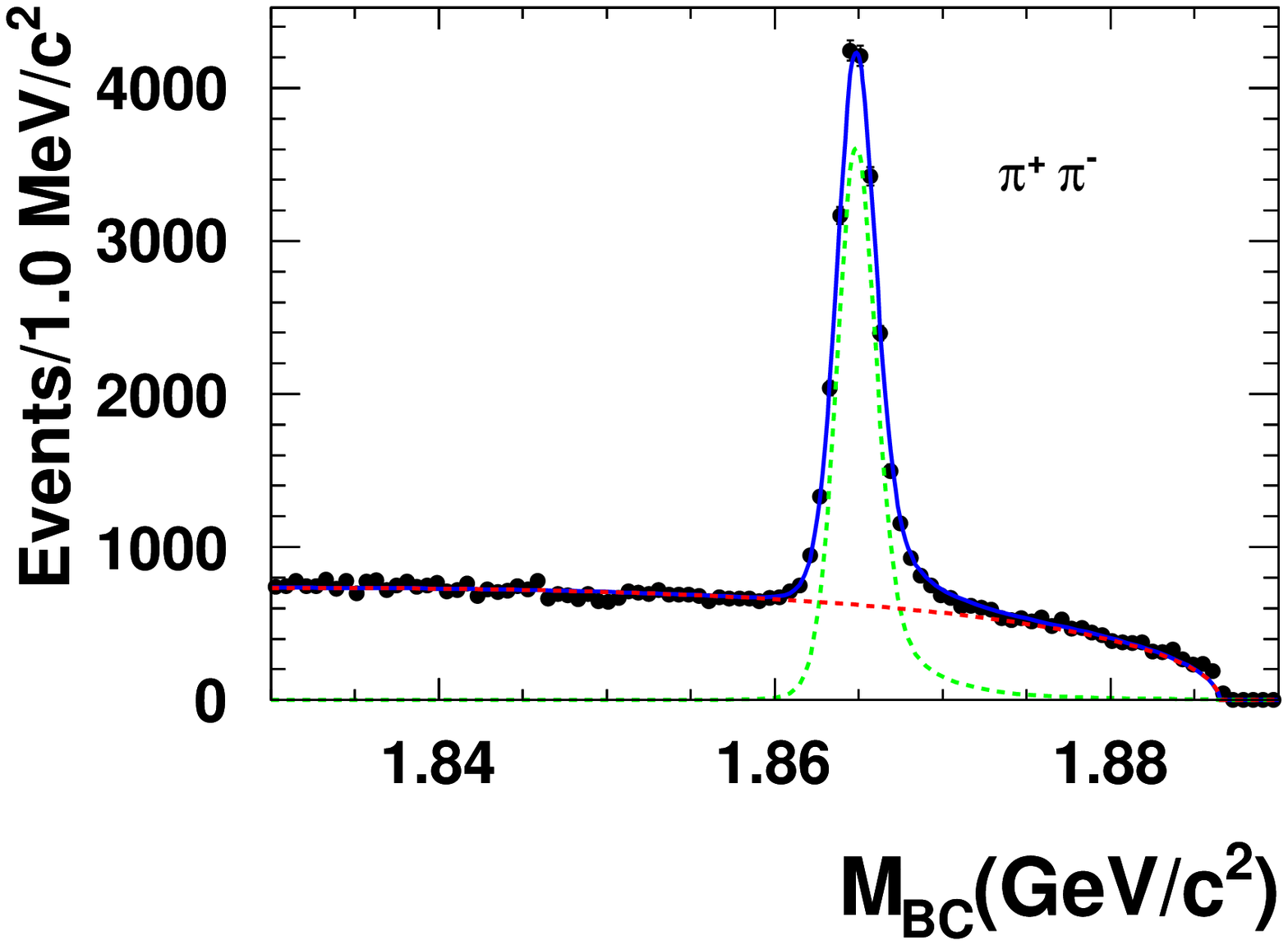}
\includegraphics[width=4.cm]{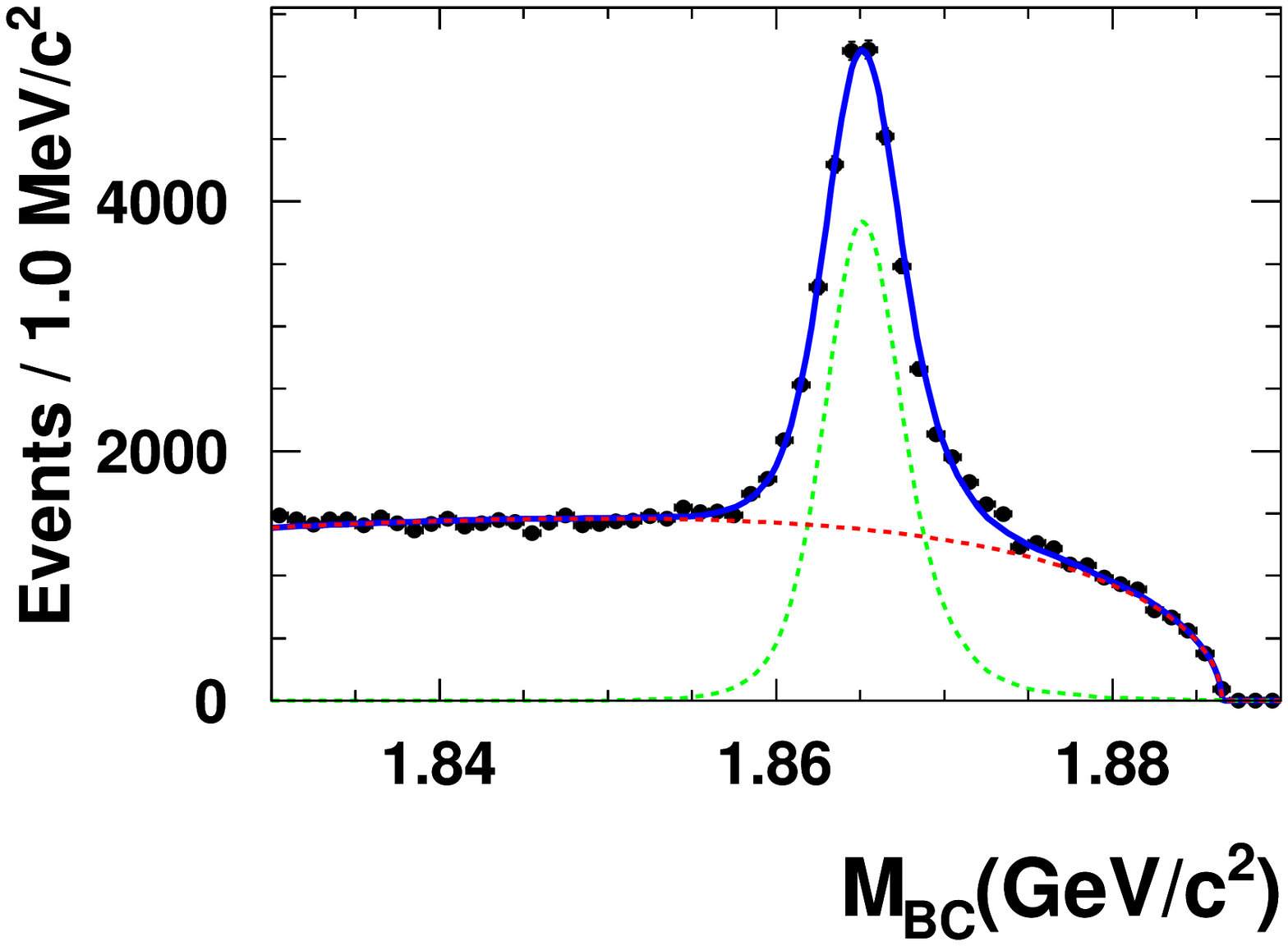}
\includegraphics[width=4.cm]{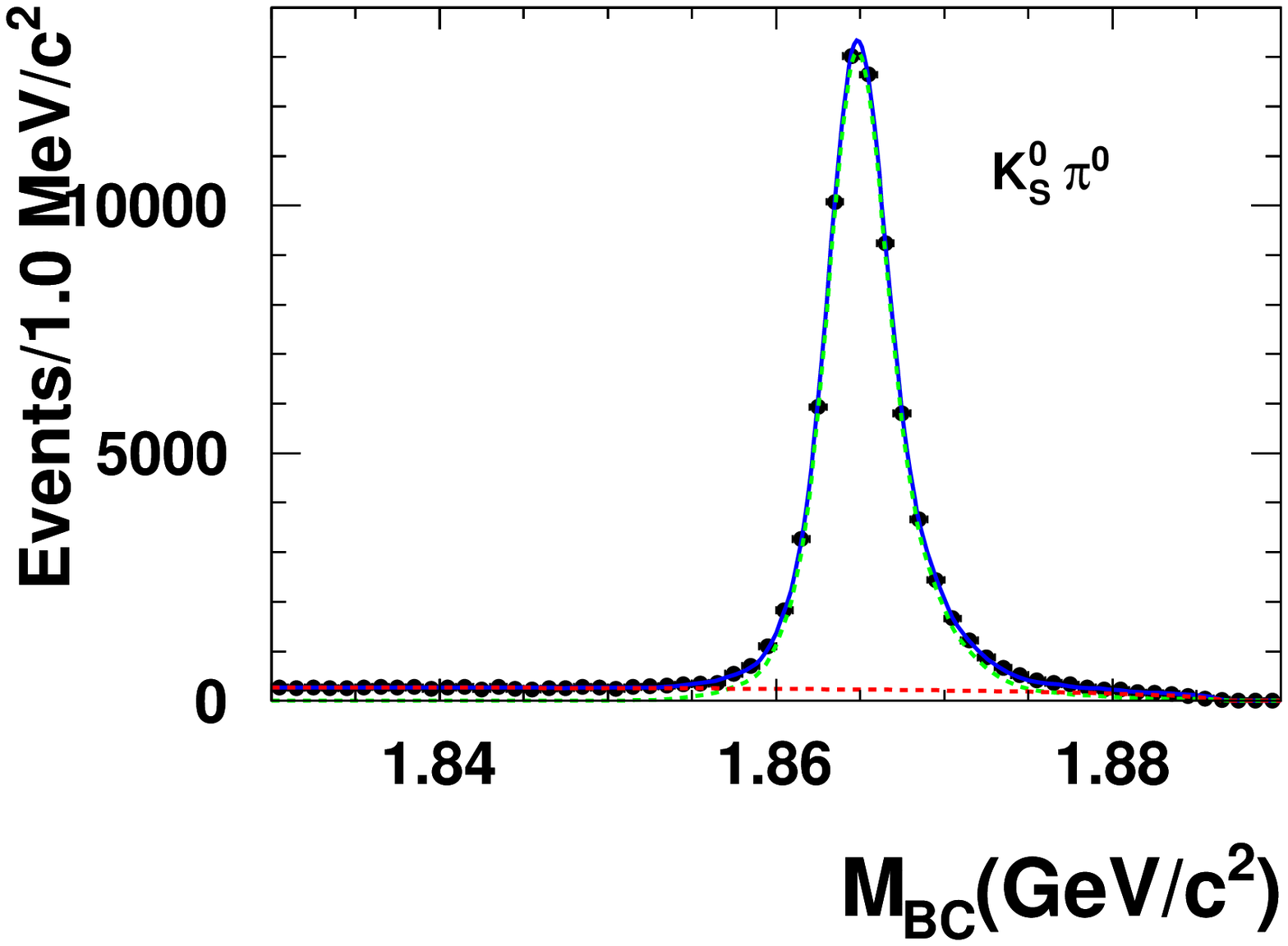}
\includegraphics[width=4.cm]{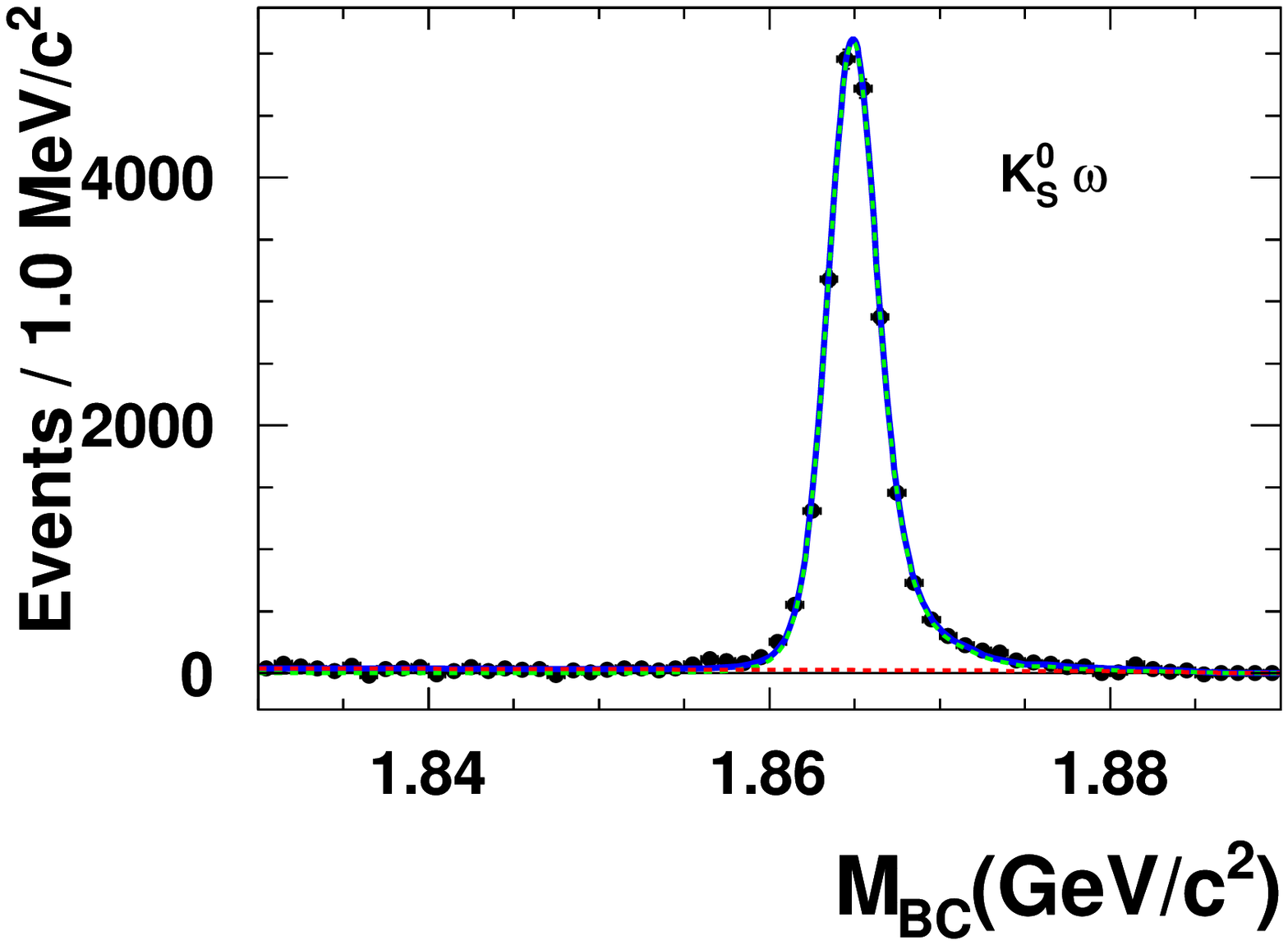}
\includegraphics[width=4.cm]{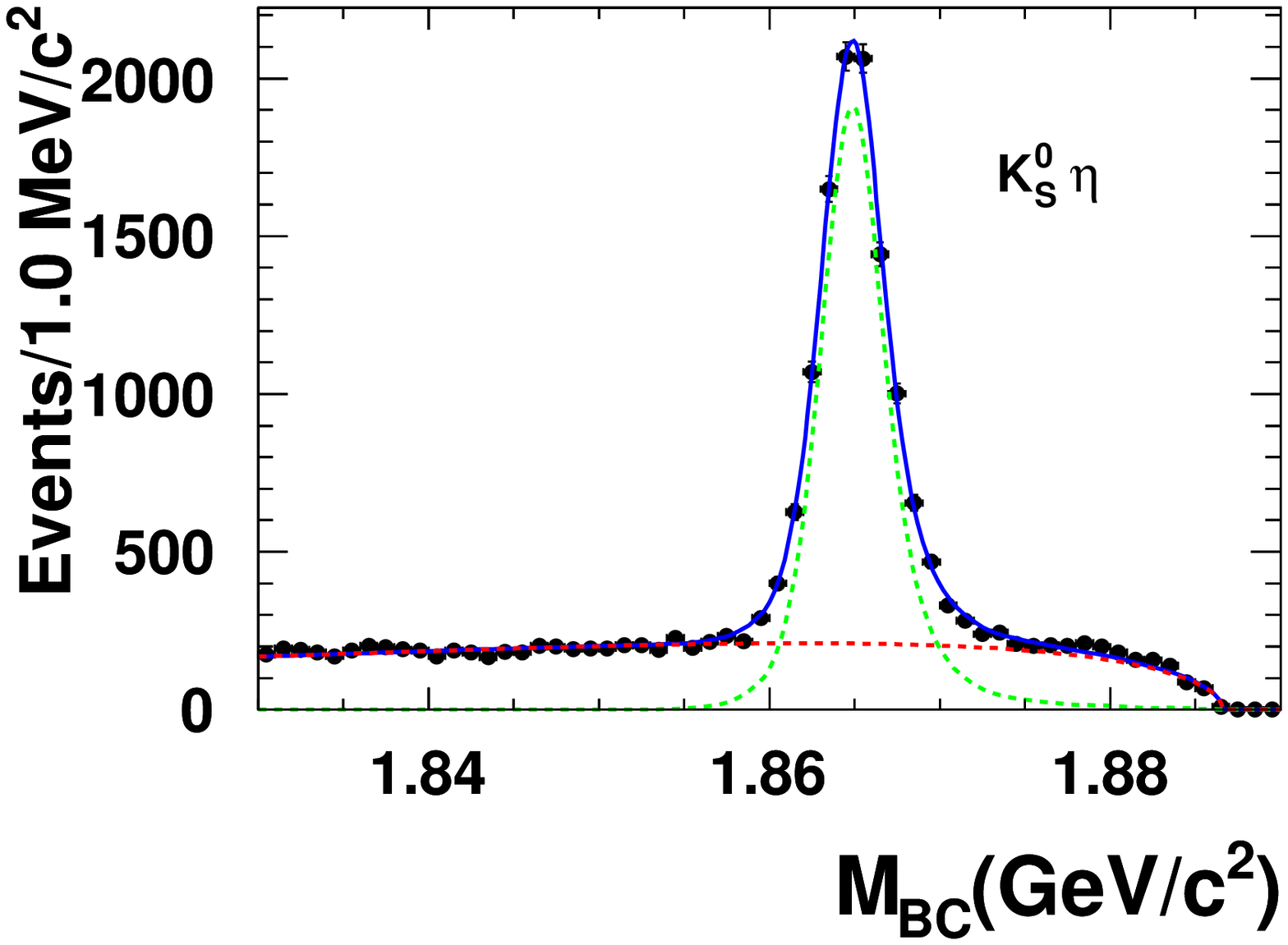}
\caption{$\mbc$ distributions and fits to data.}
\label{fig:mbcfits}
\end{center}
\end{figure}

\begin{figure}[h]
\begin{center}
\includegraphics[width=4.cm]{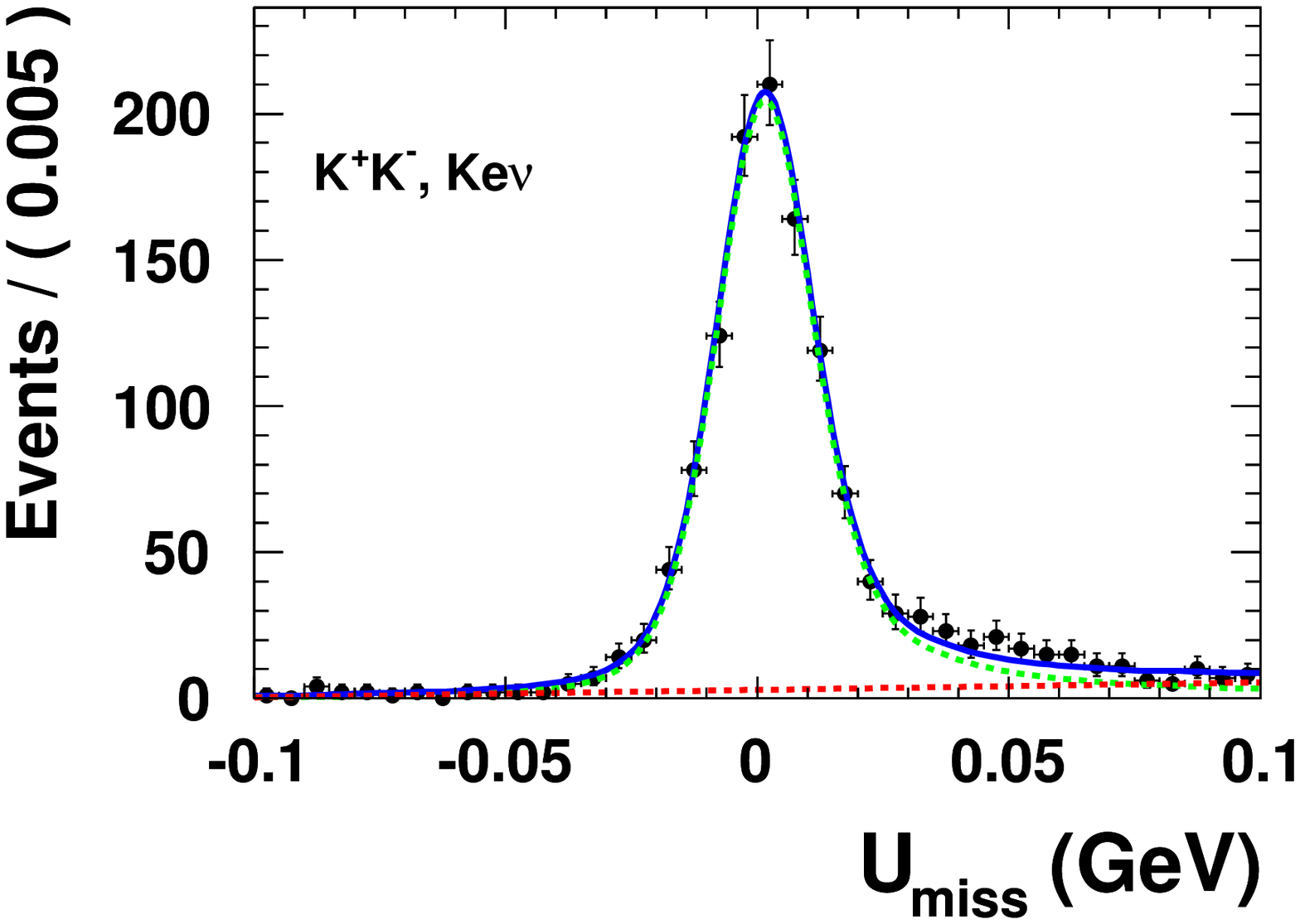}
\includegraphics[width=4.cm]{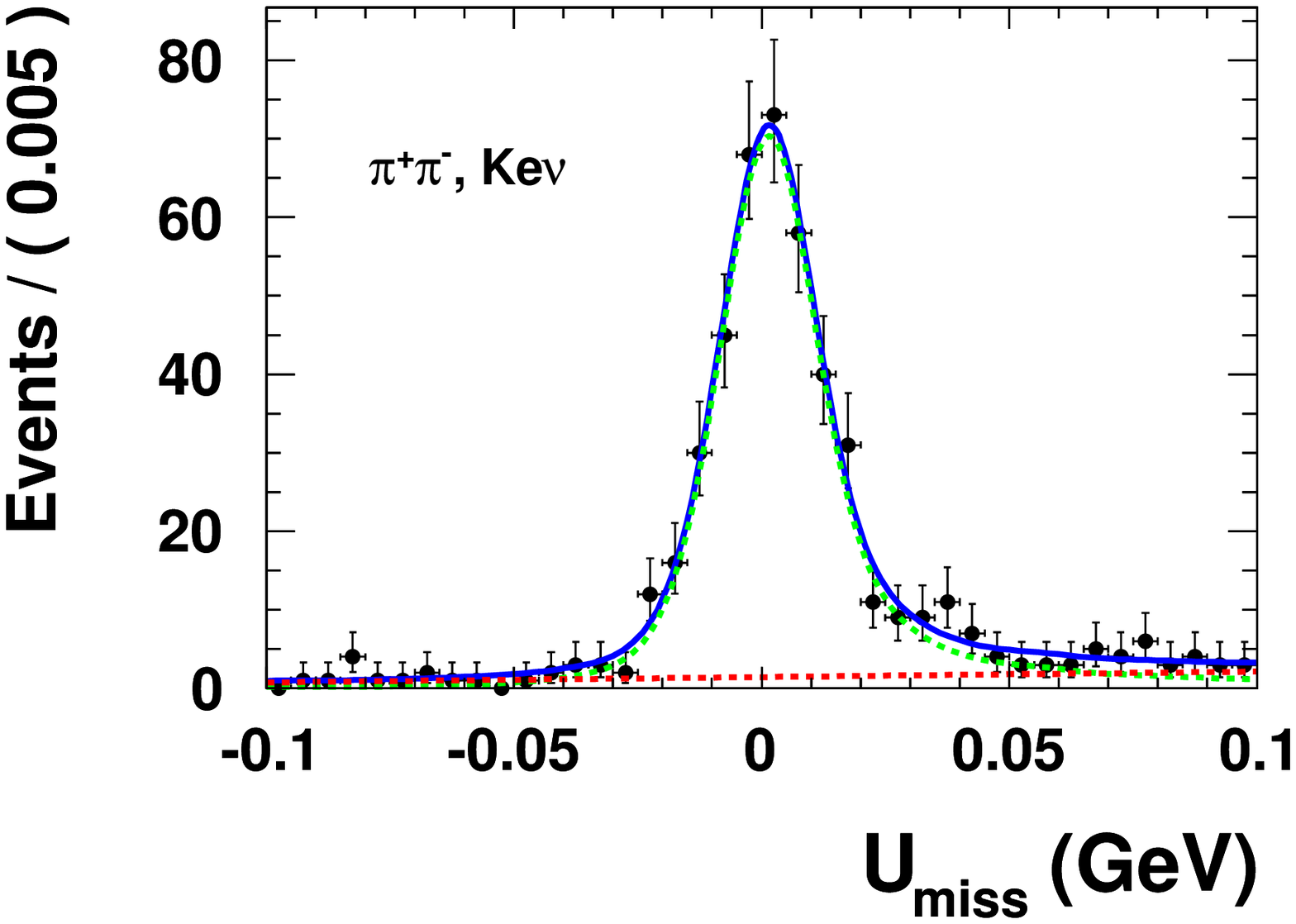}
\includegraphics[width=4.cm]{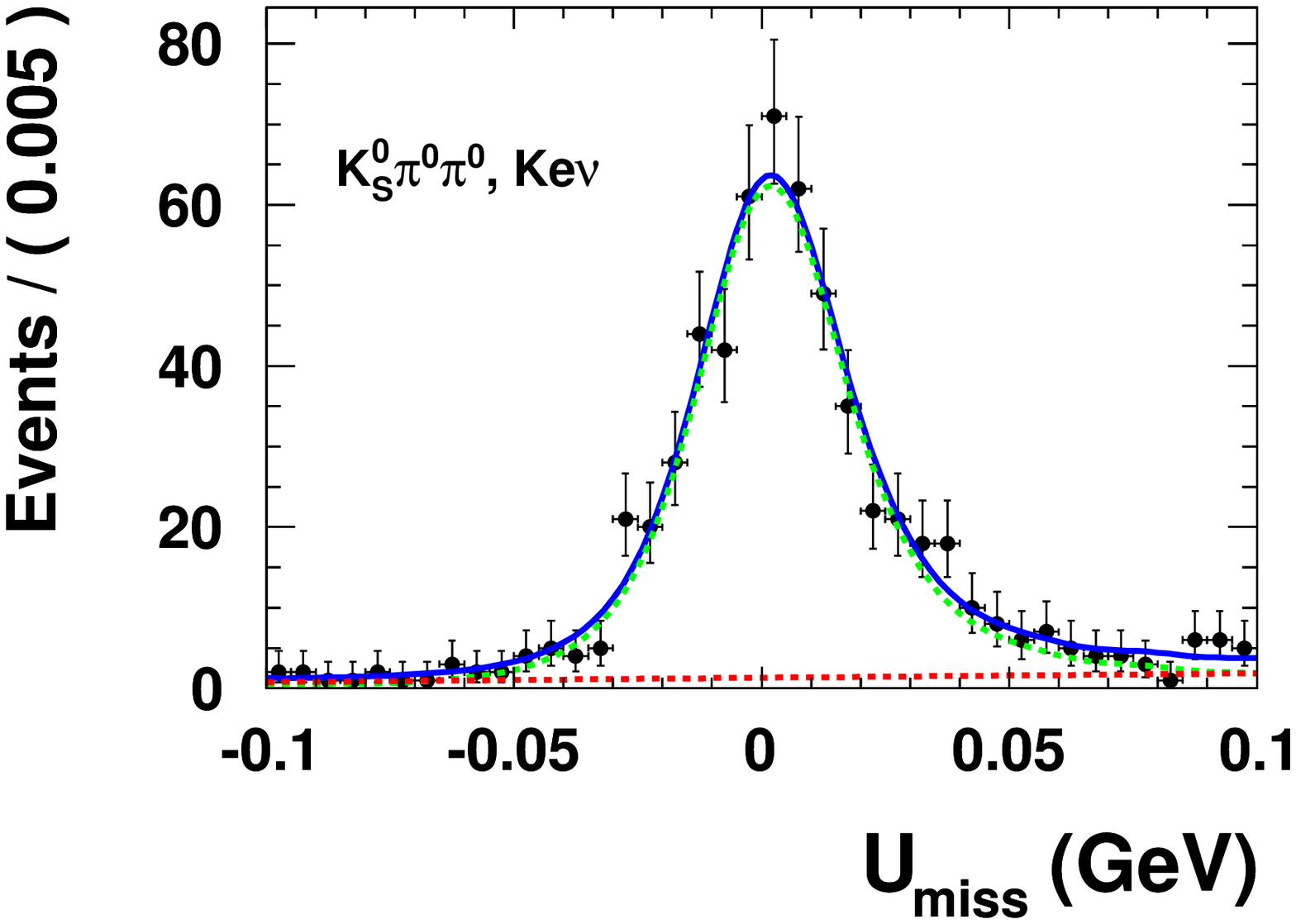}
\includegraphics[width=4.cm]{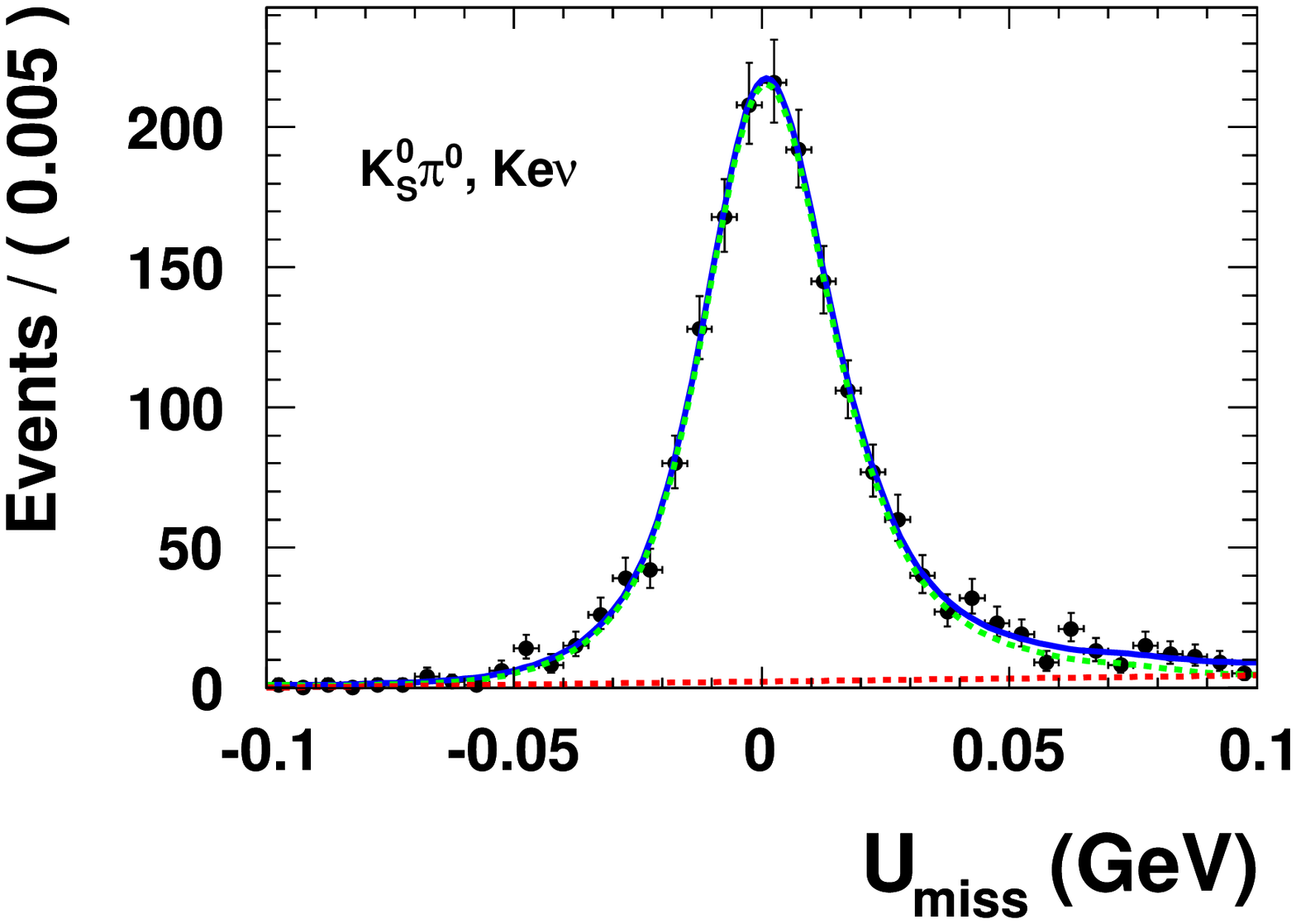}
\includegraphics[width=4.cm]{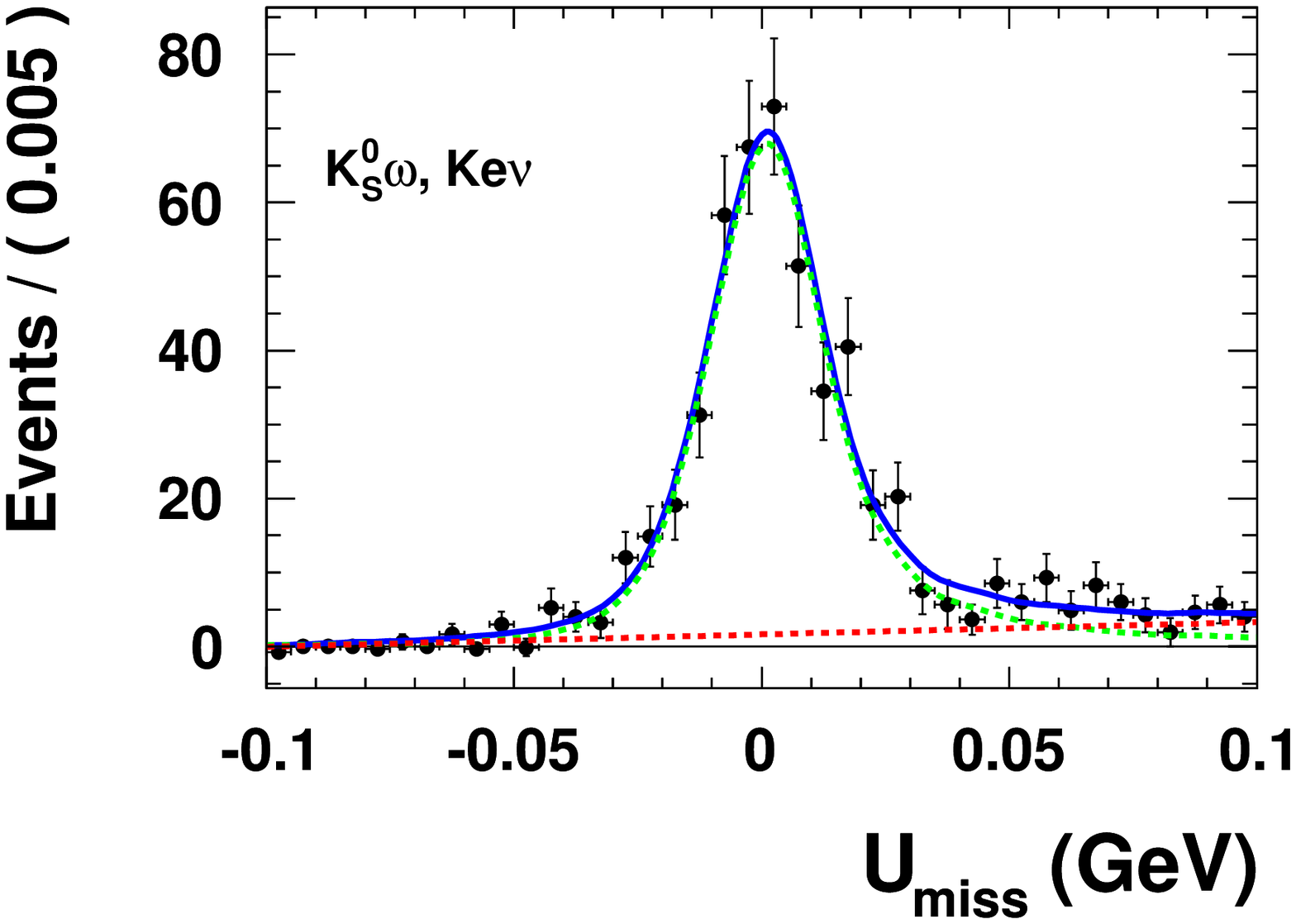}
\includegraphics[width=4.cm]{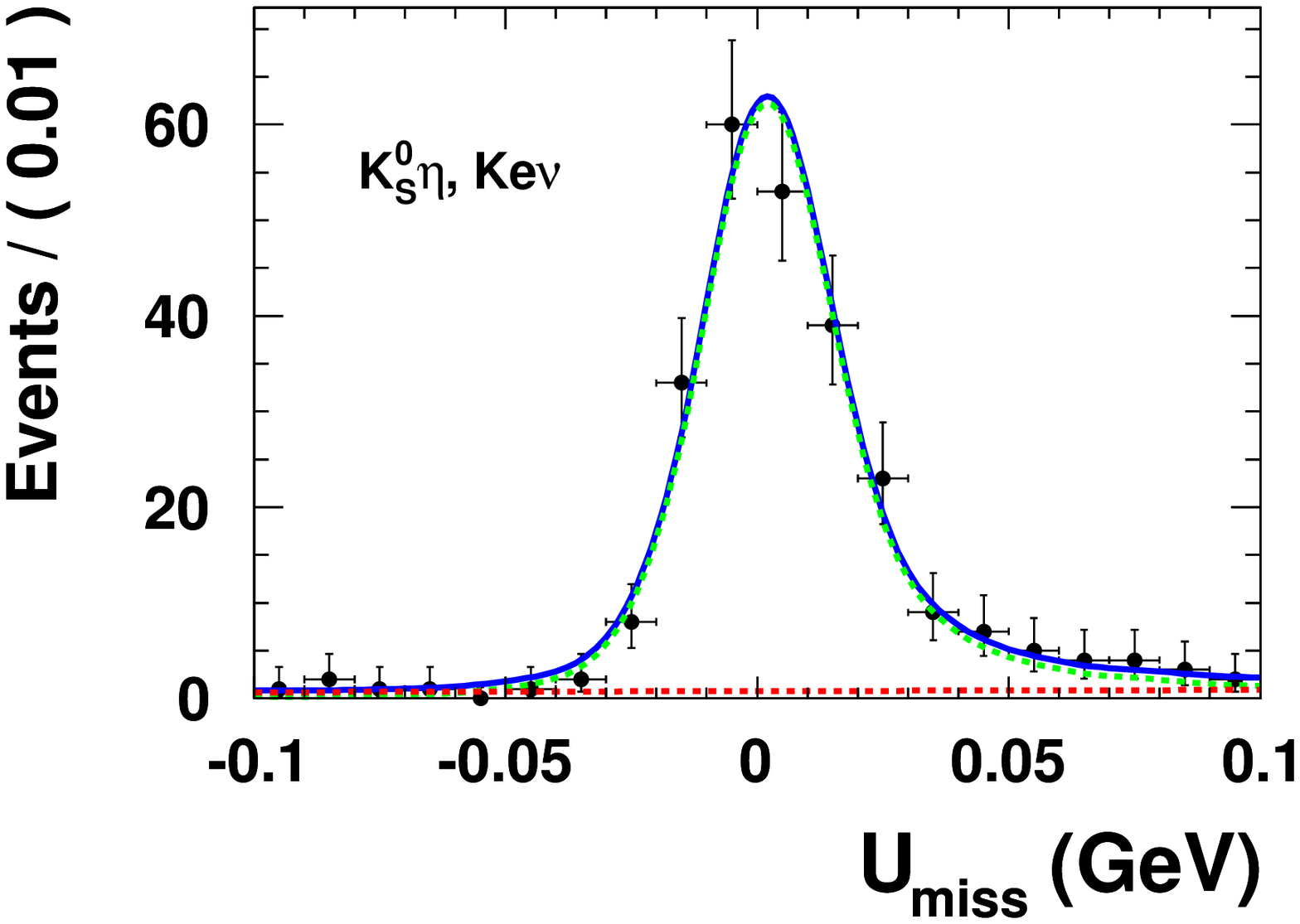}
\includegraphics[width=4.cm]{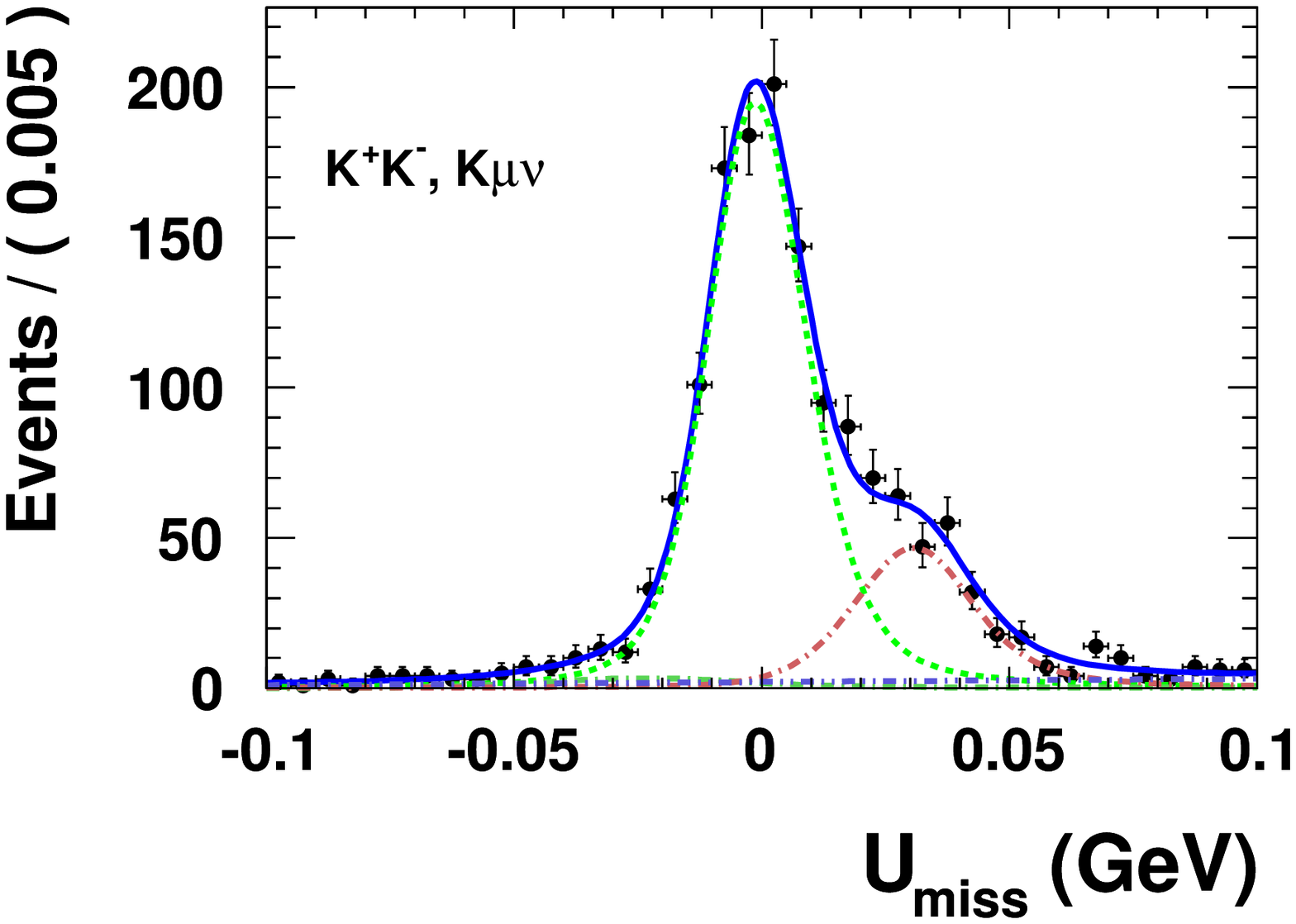}
\includegraphics[width=4.cm]{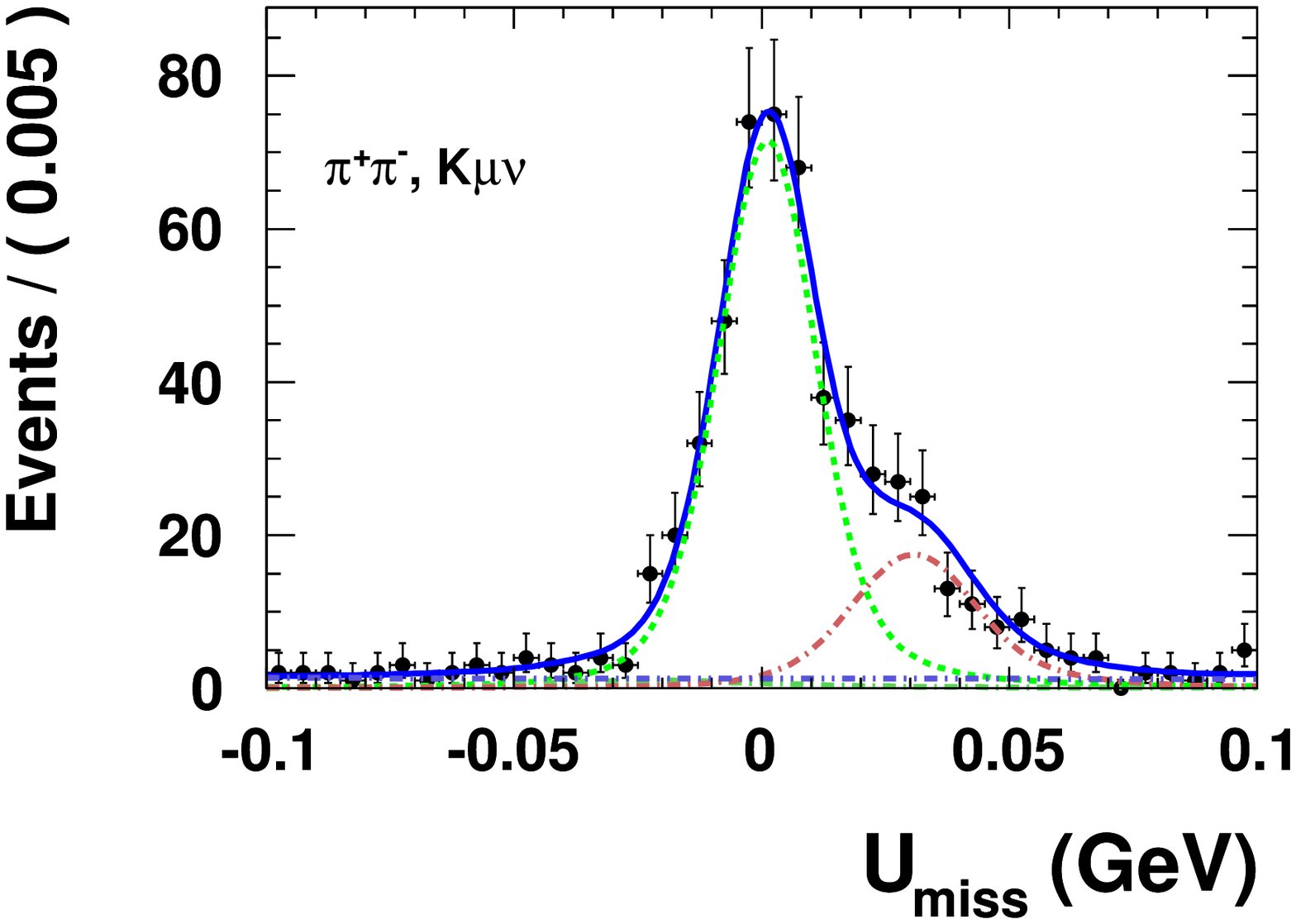}
\includegraphics[width=4.cm]{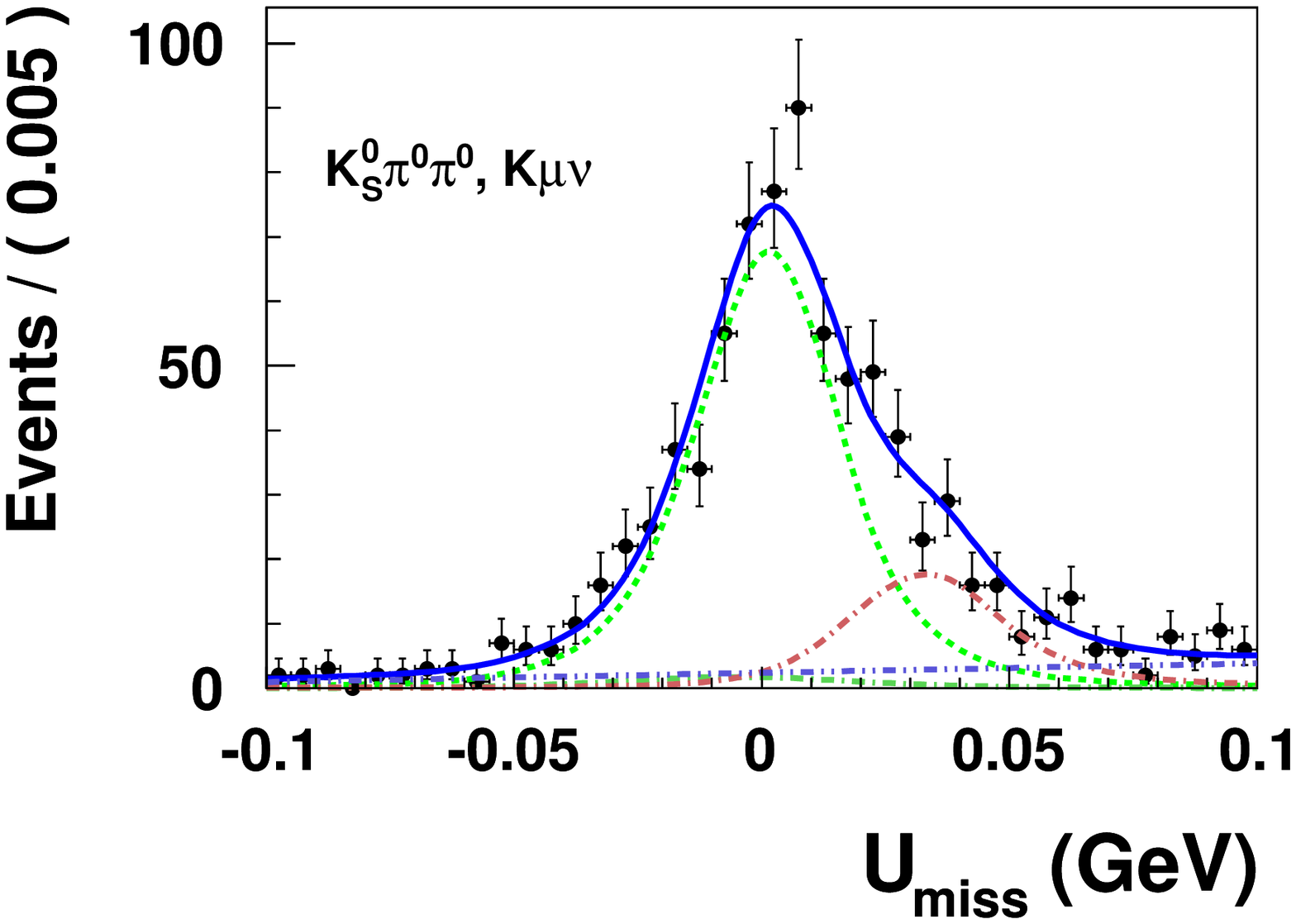}
\includegraphics[width=4.cm]{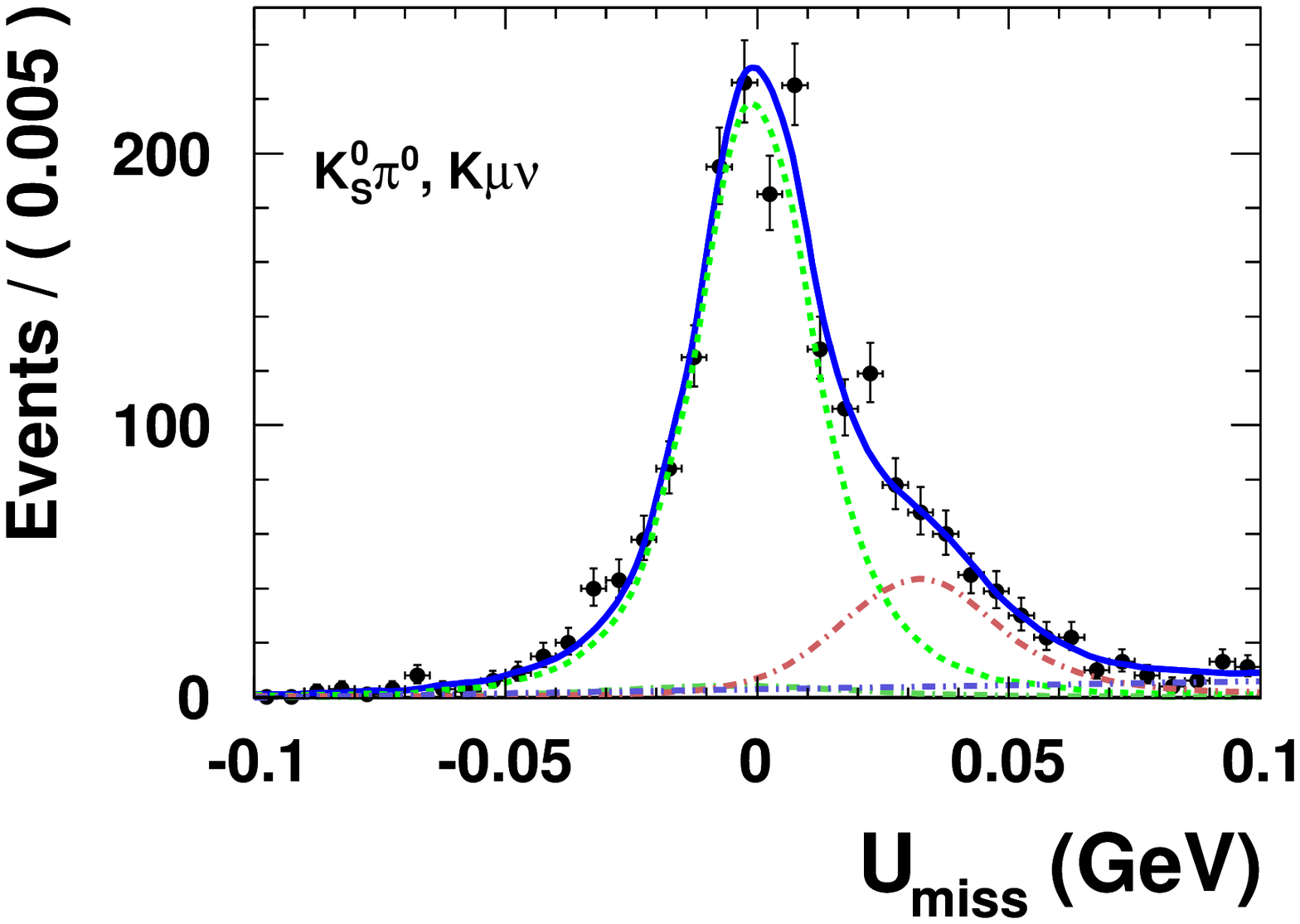}
\includegraphics[width=4.cm]{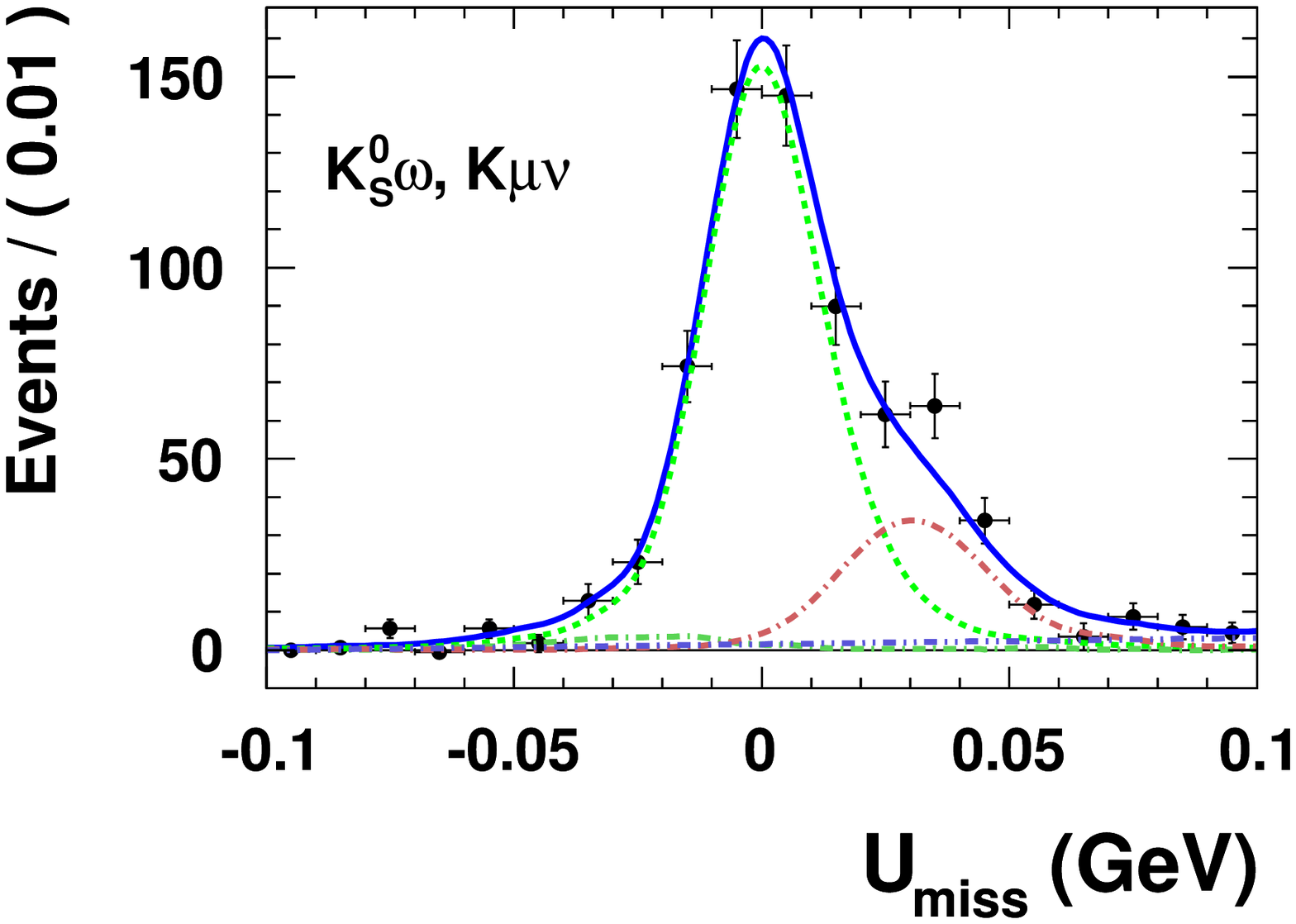}
\includegraphics[width=4.cm]{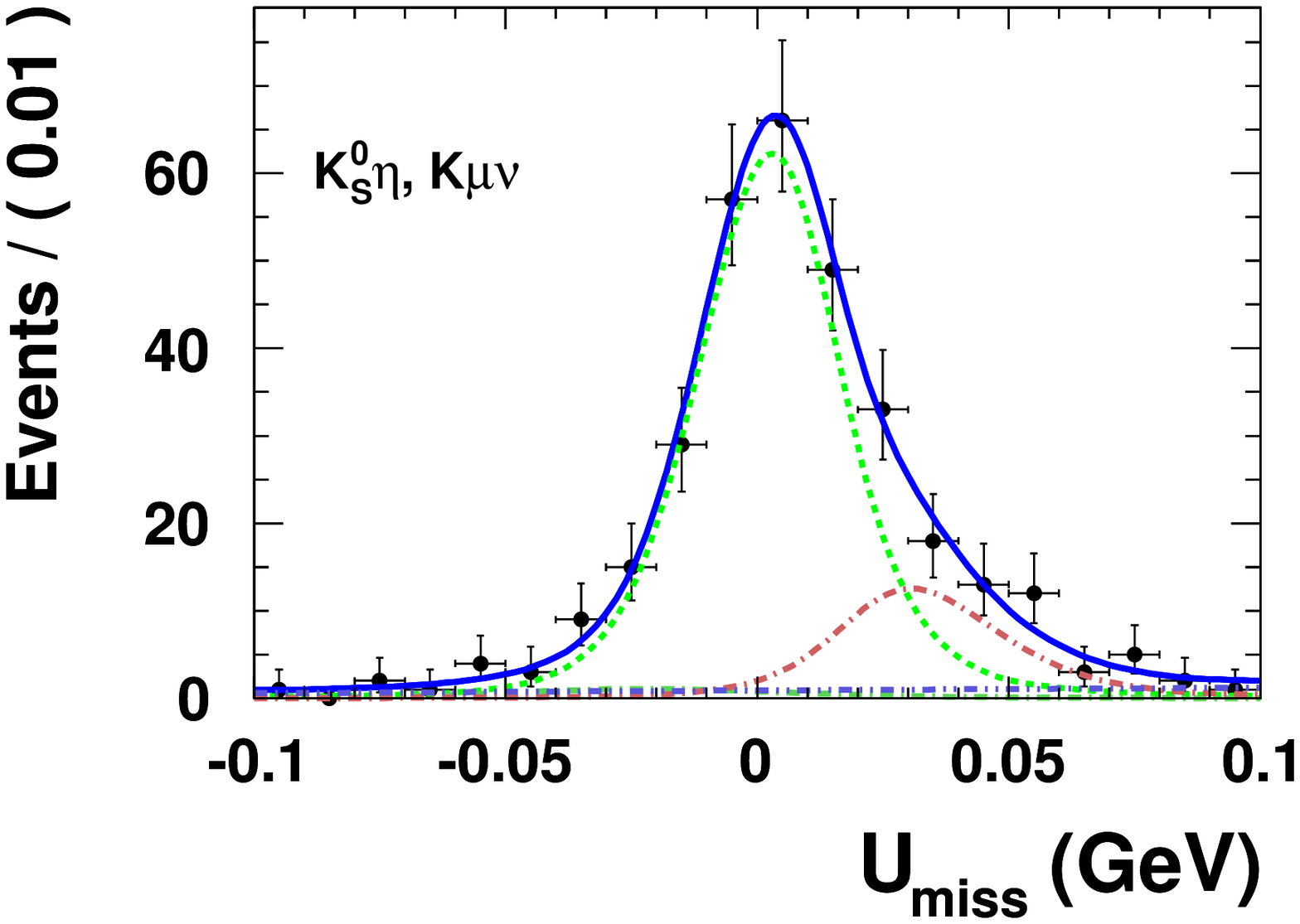}
\caption{Fits to $U_{\rm miss}$ distributions in data for $CP$-tagged $Ke\nu$ and $K\mu\nu$ modes.}
\label{fig:ufit}
\end{center}
\end{figure}

$CP$-tag modes in Tab.~\ref{table:modes} are used in this analysis.  Similar to the analysis of $\strph$, ST yields are estimated by fits to the $\mbc$ distributions, as shown in Fig.~\ref{fig:mbcfits}. Semileptonic decays of $D\to K e \nu$ and $D\to K \mu \nu$ are selected with respect to the $CP$-tagged $D$ candidates in ST events. Due to the undetectable neutrino in the final states, variable $U_{\rm miss}$ is used to distinguish the signals of semileptonic decays from backgrounds. The definition is given as
$$U_{\rm miss}\equiv E_{\rm miss}-|\vec{p}_{\rm miss}|,$$
$$E_{\rm miss} \equiv E_0-E_{K}-E_{l}, ~~~ \vec{p}_{\rm miss} \equiv -[ \vec{p}_K + \vec{p}_l + \hat{p}_{\rm ST}\sqrt{E_0^2-m_D^2}].$$
Here, $E_{K/l}$ ($\vec{p}_{K/l}$) is the energy (three-momentum) of $K^\pm$ or lepton $l^\mp$, $\hat{p}_{\rm ST}$ is the unit vector in the reconstructed direction of the $CP$-tagged $D$ and $m_D$ is the nominal $\dzero$ mass. The  $U_{\rm miss}$  distributions are plotted in Fig.~\ref{fig:ufit} for $D\to K e \nu$ and $D\to K \mu \nu$ modes.

In fits of the DT $K e \nu$ modes, signal shape is modeled using MC shape convoluted with an asymmetric Gaussian and backgrounds are described with a 1st-order polynomial function.
In fits of the DT $K \mu \nu$ modes, signal shape is modeled using MC shape convoluted with an asymmetric Gaussian. Backgrounds of $Ke\nu$ are modeled using MC shape and their relative rate to the signals are fixed. Shape of $K\pi\pi^{0}$ backgrounds are taken from MC simulations with convolution of a smearing Gaussian function; parameters of the smearing function are fixed according to fits to the control sample of $D\to K\pi\pi^{0}$ events. Size of $K\pi\pi^{0}$ backgrounds are fixed by scaling the number of $K\pi\pi^{0}$ events in the control sample to the number in the signal region according to the ratio estimated from MC simulations. Other backgrounds are described with a 1st-order polynomial function.

Finally, we obtain the preliminary result as
$$y_{CP}=-1.6\%\pm 1.3\%(\rm stat.)\pm 0.6\%(\rm syst.).$$
The result is compatible with the previous measurements~\cite{HFAG}. This is the most precise measurement of $\ycp$ based on $\dzero\dbarzero$ threshold productions. However, its precision is still statistically limited.

\section{Summary}

In  this paper, the preliminary BESIII results of the strong phase difference $\cos\delta_{K\pi}$ in $D\to K\pi$ decays and  the mixing parameter $\ycp$ are reported. The measurements were carried out based on the quantum-correlated technique in studying the process of $D^0\overline{D}{}^0$ pair productions of 2.92\,fb$^{-1}$ $e^+e^-$ collision data collected with the BESIII detector at $\sqrt{s}$ = 3.773\gev. The preliminary results are given as $$\cos\strph = 1.03\pm0.12\pm0.04\pm0.01$$ and $$y_{CP}=-1.6\%\pm 1.3\%\pm 0.6\%.$$
Among them, the result of $\cos\strph$ is the most accurate to date. In the future, global fits can be implemented  in order to best exploit BESIII data in the quantum-coherence productions~\cite{guan}.

\Acknowledgements

The BESIII collaboration thanks the staff of BEPCII and the computing center for their strong support.
This work is supported in part by Joint Funds of National Natural Science Foundation of China (11079008)
and Natural Science Foundation of China (11275266).


\begin{thebibliography}{99}

\bibitem{GIM}
S.~L.~Glashow, J.~Illiopoulos, and L.~Maiani, Phys.\ Rev.\ D {\bf 2} (1970) 1285;
R.~L.~Kingsley, S.~B.~Treiman, F.~Wilczek, and A.~Zee, Phys.\ Rev.\ D {\bf 11} (1975) 1919.

\bibitem{CKM}
M.~Kobayashi and T.~Maskawa, Prog.\ Theor.\ Phys.\ {\bf 49} (1973) 652.

\bibitem{Bigi}
S.~Bianco, F.~L.~Fabbri, D.~Benson and I.~Bigi, Riv.\ Nuovo Cim.\ {\bf 26N7} (2003) 1.

\bibitem{Xing:1996pn}
Z.-Z.~Xing, Phys.\ Rev.\ D {\bf 55} (1997) 196.

\bibitem{PDG}
J.~Beringer {\it et al.} [Particle Data Group], Phys.\ Rev.\ D {\bf 86} (2012) 010001.

\bibitem{yp1}
R.~Aaij {\it et al.} [LHCb Collaboration],  arXiv:1309.6534 [hep-ex];
R.~Aaij {\it et al.} [LHCb Collaboration], Phys.\ Rev.\ Lett.\ {\bf 110} (2013) 101802;
T.~Aaltonen {\it et al.} [CDF Collaboration], Phys.\ Rev.\ Lett.\ {\bf 100} (2008) 121802;
B.~Aubert {\it et al.} [BABAR Collaboration], Phys.\ Rev.\ Lett.\ {\bf 98} (2007) 211802.


\bibitem{Cheng:2007uj}
  X.-D.~Cheng {\it et al.},
  Phys.\ Rev.\ D {\bf 75} (2007) 094019.

\bibitem{:2009vd}
  M.~Ablikim {\it et al.}  [BESIII Collaboration],
  Nucl.\ Instrum.\ Meth.\  A {\bf 614}  (2010) 345.

\bibitem{Asner:2008ft}
  D.~M.~Asner {\it et al.}  [CLEO Collaboration],
  Phys.\ Rev.\ D {\bf 78} (2008) 012001.

\bibitem{CLEO-c2}
  D.~M.~Asner {\it et al.}  [CLEO Collaboration],
  Phys.\ Rev.\ D {\bf 86} (2012) 112001.


\bibitem{Argus}
  H.~Albrecht {\it et al.}  [ARGUS Collaboration],
  Phys.\ Lett.\ B {\bf 241} (1990) 278.

\bibitem{HFAG}
 Heavy Flavor Averaging Group: {\footnotesize{\url{http://www.slac.stanford.edu/xorg/hfag/charm/}}}.

\bibitem{guan}
  Y.~Guan, X.-R.~Lu, Y.~Zheng and Y.-S.~Zhu,
  Chinese Phys.\  C {\bf 37} (2013) 106201
  [arXiv:1304.6170 [hep-ex]].

\end{thebibliography}
\end{document}